\documentclass[reprint,a4paper,aps,prl,amsmath,amsfonts,showpacs,floatfix,nofootinbib]{revtex4-1}
\usepackage{bm}                      
\usepackage{graphicx}                
\usepackage{dcolumn}                 
\usepackage[table]{xcolor}
\usepackage{amsmath}
\usepackage{times}

\allowdisplaybreaks

\newcommand*{\centt}[1]{\multicolumn{1}{c}{#1}}
\newcommand*{\cent}[1]{\multicolumn{1}{c}{$#1$}}
\newcolumntype{x}[1]{D{.}{.}{#1}}

\newcommand{\icm}{\text{cm}^{-1}}

\newcommand{\m}[1]{m_{\mathrm{#1}}}

\begin{document}
\preprint{Version 0.9}

\title{Dissociation Energy of Molecular Hydrogen Isotopologues}
\author{Mariusz Puchalski}
\affiliation{Faculty of Chemistry, Adam Mickiewicz University, Uniwersytetu Pozna\'nskiego 8, 61-614 Pozna{\'n}, Poland}

\author{Jacek Komasa}
\affiliation{Faculty of Chemistry, Adam Mickiewicz University, Uniwersytetu Pozna\'nskiego 8, 61-614 Pozna{\'n}, Poland}

\author{Anna Spyszkiewicz}
\affiliation{Faculty of Chemistry, Adam Mickiewicz University, Uniwersytetu Pozna\'nskiego 8, 61-614 Pozna{\'n}, Poland}

\author{Krzysztof Pachucki}
\affiliation{Faculty of Physics, University of Warsaw, Pasteura 5, 02-093 Warsaw, Poland}

\date{\today}

\begin{abstract}

  The nonrelativistic energy together with relativistic and quantum electrodynamic corrections for all the molecular
  hydrogen isotopologues (D$_2$, T$_2$, HD, HT, DT) were evaluated without expansion in the electron-nucleus mass ratio.
  The obtained results significantly improve the uncertainty of theoretical predictions, reaching
  a value below 1 MHz for the total dissociation energy. We observe good agreement
  with the experimental value for D$_2$ and $3\,\sigma$ discrepancy for the HD molecule,
  while no experimental values for the dissociation energy of molecules involving tritium have yet been obtained.
\end{abstract}
\maketitle

\section{Introduction}

The dissociation energy of ortho-H$_2$ has recently been measured with sub-MHz uncertainty \cite{Cheng:18,Holsch:19},
which is smaller than the contribution due the finite size of the proton. 
Therefore, for the first time the molecular hydrogen spectroscopy has become sensitive to the nuclear charge radius.  
Similar progress is expected for HD and D$_2$ systems, which have been
measured so far with 10 MHz uncertainty \cite{Liu:10,Sprecher:10}.
Moreover, very accurate measurements of several molecular transitions for tritium-containing isotopologues
have been obtained \cite{Trivikram:18,Lai:19} very recently,
which indicates the possibility of measurements of their dissociation energies in the future.

The current theoretical dissociation energy of para-H$_2$ is $36\,118.069\,632(26)$ cm$^{-1}$ 
\cite{PKCP:19}, which corresponds to 0.8 MHz of absolute uncertainty. It is in a good agreement
with the most recent experimental value of $36\,118.069\,45(31)$ cm$^{-1}$ \cite{Altmann:18},
which will soon be improved \cite{Hussels:19}.
In this work we demonstrate that similar accuracy can be reached
for all the other molecular hydrogen isotopologues: D$_2$, T$_2$, HD, HT, and DT.

The high accuracy of theoretical predictions for molecular levels can only be achieved with the approach,
based on nonrelativistic quantum electrodynamic (NRQED) theory.
According to NRQED, the total energy of an atom or a molecule can be represented by the
expansion in powers of the fine structure constant $\alpha$
\begin{equation}\label{alphaexp}
  E(\alpha) = \sum_{n=2}m\alpha^n\,E^{(n)}.
\end{equation}
In our previous work on H$_2$ \cite{PKCP:19}, the first three terms---the nonrelativistic energy $E^{(2)}$,
the relativistic correction $E^{(4)}$, and the leading QED correction $E^{(5)}$---were
calculated to a high numerical precision in direct four-body variational calculations. 
The higher order QED corrections were evaluated in the framework 
of the Born-Oppenheimer (BO) approximation, wherein the $E^{(6)}$ correction 
was evaluated in a complete way, but $E^{(7)}$ was merely estimated from the dominating terms,
which are known from the hydrogen atom \cite{Eides:01}. 
Here, the same method is applied to the heavier homonuclear isotopologues
D$_2$ and T$_2$, as well as extended further to heteronuclear systems HD, HT, and DT.
The uncertainties assigned to theoretical predictions are below 1~MHz,
which is an improvement by 1--2 orders of magnitude compared to the most accurate 
previous results.

\section{Nonrelativistic wave function}

The quality of the determination of the nonrelativistic wave function is of critical importance
for achieving the high accuracy of theoretical results. 
In the direct nonadiabatic approach, in which all particles are treated 
on an equal footing, the wave function $\Psi$ is a solution to the four-body Schr{\"o}dinger equation $H\,\Psi = E\,\Psi$
with  $E=E^{(2)}$ and the Hamiltonian
\begin{align}
H &= T+V\,, \\
T		 &=\frac{ \vec p_0^{\,2}}{2\,m_0} + \frac{ \vec p_1^{\,2}}{2\,m_1} +  \frac{\vec p_2^{\,2}}{2 \, m} + \frac{\vec p_3^{\,2}}{2\, m}\,,  \\
V &=\frac{1}{r_{01}}-\frac{1}{r_{02}}-\frac{1}{r_{03}}-\frac{1}{r_{12}}-\frac{1}{r_{13}}+\frac{1}{r_{23}}\,. 
\end{align}
The indices 0, 1 denote nuclei, and 2, 3---electrons.
In the center-of-mass frame, the wave function $\Psi$ depends only on the interparticle distances
$r_{ij}$ and is represented as 
\begin{eqnarray}
 \Psi &=& \sum_k^N c_k\, \psi_k(\vec r_0, \vec r_1, \vec r_2, \vec r_3)\,, \\
 \psi_k &=& (1 + P_{0\leftrightarrow 1})\,(1+P_{2\leftrightarrow 3})\,\phi_{\{k\}} (\vec r_0, \vec r_1, \vec r_2, \vec r_3)\,, \label{psii}
 \end{eqnarray}
where the $P_{i\leftrightarrow j}$ operator accounts for the symmetry with respect to
the exchange of nuclei (applicable to homonuclear molecules) or electrons.
Two types of explicitly correlated basis functions $\phi_{\{k\}}$ are employed
to expand the wave function---the nonadiabatic James-Coolidge (naJC) \cite{PK:16} or the
explicitly correlated Gaussian (naECG) \cite{adamowicz_99,PSKP:18} basis.

The spatial function within the naJC approach is
\begin{equation}
\phi_{\{k\}}=e^{-\alpha\,R-\beta(\zeta_2+\zeta_3)}\,
R^{k_0}\,r_{23}^{k_1}\,\eta_2^{k_2}\,\eta_3^{k_3}\,\zeta_2^{k_4}\,\zeta_3^{k_5}\label{EJCBas}
\end{equation}
where $\zeta_2=r_{02}+r_{12}$, $\eta_2=r_{02}-r_{12}$, $\zeta_3=r_{03}+r_{13}$, $\eta_3=r_{03}-r_{13}$,
and $R = r_{01}$. The $\alpha$ and $\beta$ in Eq.~(\ref{EJCBas}) denote 
nonlinear variational parameters, common for the whole set of basis functions called `sector', 
and $k_i$ are non-negative integers collectively denoted as $\{k\}$. 
If needed, two or more sectors (with different pairs of $\alpha^{(i)}$ and $\beta^{(i)}$) can be used.  
In this work, the naJC basis was employed in calculations of the nonrelativistic energy, 
which converged up to 13 significant figures.
This basis has not been used so far for relativistic calculations, as we have not yet worked out
all the integrals needed for matrix elements with relativistic operators.

The Gaussian (naECG) basis, used here in the calculations of the the relativistic and QED corrections,
is represented by spatial functions of the form
\begin{equation}
\phi_{\{k\}} =r_{01}^n\, e^{-a_{k1} r^2_{01}-a_{k2} r^2_{02}-a_{k3} r^2_{03}-a_{k4} r^2_{12}-a_{k5} r^2_{13}-a_{k6} r^2_{23}}\,. \label{phi2} 
\end{equation}
In the particular case of expectation values of certain relativistic operators, $\phi_{\{k\}}$ is modified to
\begin{align}
  \phi_{\{k\}} =&\ r_{01}^n\,\Bigl(1+\frac{r_{23}}{2}\Bigr)\nonumber \\\label{phi2cusp} &\ 
  e^{-a_{k1} r^2_{01}-a_{k2} r^2_{02}-a_{k3} r^2_{03}-a_{k4} r^2_{12}-a_{k5} r^2_{13}-a_{k6} r^2_{23}}\,,  
\end{align}
which ensures that the nonrelativistic wave function exactly satisfies the electron-electron
cusp condition \cite{kato}. Namely, due to the electron-electron Coulomb interaction,
the exact wave function $\Psi(r_{23})$ must behave for small $r_{23}$ as
$\Psi(r_{23})\approx \Psi(0)\,(1+r_{23}/2)$ which is authomatically satisfied in the above basis.
The internuclear $r_{01}^n$ prefactor enables proper representation 
of the vibrational part of the wave function. The powers $n$ of this coordinate 
are restricted to even integers within the range $0-80$ and are generated following 
the log-normal distribution. The nonlinear $a_{kl}$ parameters are determined variationally
in an extensive optimization process. 
The naECG wave function $\Psi$ has been optimized for a sequence of growing basis set sizes
to observe the convergence of the nonrelativistic energy.
This convergence is presented in Table~\ref{TE2conv} and compared with the results
of naJC calculations used here as a benchmark, because they are by far the most accurate ones
in the literature. As can be inferred from this table,
the naECG nonrelativistic energy is converged to at least 10 significant figures.

\begin{table*}[hbt]
\caption{Convergence of the nonrelativistic energy $E^{(2)}$ (in a.u.) with the increasing size $N$ of the naECG basis set
in comparison with the benchmark values from nonadiabatic James-Coolidge (naJC) wave function. The following CODATA 2018
\cite{CODATA_18} mass ratios were used in these calculations:
$\m{p}/m = 1\,836.152\,673\,43(11)$, $\m{d}/m = 3\,670.482\,967\,88(13)$,  $\m{t}/m = 5\,496.921\,535\,73(27)$.
}
\label{TE2conv}
\begin{ruledtabular}
\begin{tabular}{cx{3.17}x{3.17}x{3.17}x{3.17}x{3.17}}
\cent{N} & \centt{D$_2$} & \centt{T$_2$} & \centt{HD} & \centt{HT} & \centt{DT} \\
\hline\\[-1.5ex]
128      & -1.167\,167\,911\,358 & -1.168\,534\,104\,823 & -1.165\,470\,991\,485 & -1.166\,000\,790\,842 & -1.167\,817\,701\,507 \\
256      & -1.167\,168\,756\,439 & -1.168\,535\,448\,080 & -1.165\,471\,628\,967 & -1.166\,001\,763\,875 & -1.167\,819\,489\,839 \\
512      & -1.167\,168\,805\,491 & -1.168\,535\,668\,007 & -1.165\,471\,916\,621 & -1.166\,002\,029\,805 & -1.167\,819\,626\,122 \\
1024     & -1.167\,168\,808\,953 & -1.168\,535\,674\,847 & -1.165\,471\,923\,256 & -1.166\,002\,036\,615 & -1.167\,819\,671\,730 \\
2048     & -1.167\,168\,809\,201 & -1.168\,535\,675\,524 & -1.165\,471\,923\,906 & -1.166\,002\,037\,196 & -1.167\,819\,673\,214 \\[1ex]
naJC     & -1.167\,168\,809\,284\,10(5) & -1.168\,535\,675\,732\,90(8) & -1.165\,471\,923\,963\,66(5) & -1.166\,002\,037\,328\,67(6) & -1.167\,819\,673\,436\,73(5) \\
\end{tabular}                            
\end{ruledtabular}   
\end{table*}

\section{The relativistic correction}
The relativistic correction can be expressed in terms of the expectation value
\begin{equation}
E^{(4)} = \langle \Psi | H_\mathrm{rel}| \Psi \rangle
\label{ERdirect}
\end{equation}
of the Breit-Pauli Hamiltonian ($m=1$)
\begin{eqnarray}
H_\mathrm{rel} &=& -\frac{1}{8} (p_2^4 + p_3^4)
+\frac{\pi}{2} \sum_{x,a} \bigg(1 + \frac{\delta_s^x}{m_x^2}\bigg)\delta^3(r_{xa})   \nonumber \\
&& + \pi\,\delta^3(r_{23}) -\frac{1}{2}\,p_2^i\,\biggl(\frac{\delta^{ij}}{r_{23}}+\frac{r_{23}^i\,r_{23}^j}{r_{23}^3}\biggr)\,p_3^j  \nonumber \\
&& + \frac{1}{2}  \sum_{x,a} \frac{1}{m_x} p_x^i\,\biggl(\frac{\delta^{ij}}{r_{xa}}+\frac{r_{xa}^i\,r_{xa}^j}{r_{xa}^3}\biggr)\,p_a^j \nonumber \\
&& - \frac{1}{2}  \frac{1}{m_0\,m_1} p_0^i\,\biggl(\frac{\delta^{ij}}{r_{01}}+\frac{r_{01}^i\,r_{01}^j}{r_{01}^3}\biggr)\,p_1^j \\\nonumber
\end{eqnarray}
where index $x$ goes over nuclei and $a$ over electrons. The coefficient $\delta_s^x = 0$ 
for the nuclear spin  $s=0$ or 1, and $\delta_s^x = 1$ for $s=1/2$ \cite{muonic_d}. 
In the above formulas, we have omitted all the electron spin-dependent terms
because they vanish for the ground electronic state of $^1\Sigma_g^+$ symmetry.
Moreover, we have omitted also the $p_x^4/(8\,m_x^3)$ and $\delta^3(r_{01})$
terms because their numerical values are smaller than
the uncertainty of the whole relativistic correction. 

The result for relativistic correction to the dissociation energy $D_0$ is shown in Table~\ref{TQ}.
$D_0$ differs from the expectation values of $H_{\rm rel}$ by subtraction
of the corresponding energy of separated atoms,
\begin{eqnarray}\label{ErelH}
  E^{(4)}_x &=& 
  -\frac{1}{8}\, + \frac{1}{4}\,\bigg(\frac{1}{m_x}\bigg)^2 + O\Big(\frac{1}{m_x}\Big)^3,
\end{eqnarray}
and the overall sign.
It is worth noting that no term proportional to $1/m_x$ is present in the above formula,
so the relativistic recoil correction for separated atoms is of higher order in the mass ratio.
Thanks to the regularization of the relativistic operators, which we performed in Ref.~\cite{PKCP:19},
and the application of the variational wave function~(\ref{phi2cusp}),
the total relativistic contribution has a very good convergence with the size of the basis set,
and the extrapolated values are accurate to at least six digits (see Tab.~\ref{TQ}).
\begin{table*}[!ht]
\caption{Convergence of relativistic correction to the dissociation energy $D_0$ (in $\icm$) 
  with the increasing size $N$ of the naECG basis set.}
\label{TQ}
\begin{ruledtabular}
\begin{tabular}{c@{\extracolsep{\fill}}x{2.9}x{2.9}x{2.9}x{2.9}x{2.9}}
\cent{N} & \centt{D$_2$} & \centt{T$_2$} & \centt{HD} & \centt{HT} & \centt{DT} \\
\hline\\[-1.5ex]
128               & -0.528\,337\,669   & -0.527\,017\,169 &  -0.529\,979\,01   & -0.529\,443\,386   & -0.527\,841\,985 \\
256               & -0.528\,218\,423   & -0.526\,738\,994 &  -0.529\,910\,95   & -0.529\,374\,084   & -0.527\,577\,232 \\
512               & -0.528\,201\,146   & -0.526\,756\,712 &  -0.529\,883\,50   & -0.529\,372\,726   & -0.527\,532\,246 \\
1024              & -0.528\,205\,416   & -0.526\,750\,343 &  -0.529\,886\,61   & -0.529\,378\,527   & -0.527\,524\,975 \\
2048              & -0.528\,205\,935   & -0.526\,750\,223 &  -0.529\,887\,30   & -0.529\,378\,110   & -0.527\,523\,876 \\[1ex]	                     
$\infty$          & -0.528\,206\,05(9) & -0.526\,750\,0(2) &  -0.529\,887\,5(2) & -0.529\,377\,9(2) & -0.527\,523\,6(3) \\
\end{tabular}                            
\end{ruledtabular}   
\end{table*}

\section{The leading QED correction}
The formula for the leading quantum electrodynamic correction $E^{(5)}$ for H$_2$ was obtained 
in Ref.~\cite{PKCP:19}.
However, the nonlogarithmic $(m/m_x)^2$ terms are unknown in the case of nuclei with
spin $s\neq 1/2$. Because their numerical contribution 
is negligibly small, such terms are absent in the formula employed here
\begin{widetext}
\begin{align}
  E^{(5)} =&\ -\frac{2{\cal D}}{3\,\pi} \ln k_0
  -\frac{7}{6\,\pi} \biggl\langle \frac{1}{r_{23}^3}
  + \sum_{a,x}\frac{m}{m_x}\frac{1}{r_{ax}^3}\biggr\rangle_{\!\epsilon}
  +\frac{4}{3}\,\sum_{a,x}\biggl\{
\biggl(1 + \frac{m}{4\,m_x} + \frac{m^2}{m_x^2}\biggr)\,\ln\bigl(\alpha^{-2}\bigr)+
 \frac{19}{30} +\frac{m}{m_x}\,\frac{31}{6}
\nonumber \\&\
+\frac{m^2}{m_x^2}\,\ln\biggl(\frac{m_x}{m}\biggr)
\biggr\}\big\langle \delta^3(r_{ax}) \big\rangle
+\left(\frac{164}{15}+\frac{14}{3}\,\ln\alpha\right)\,\big\langle\delta^3(r_{23})\big\rangle
- E^{(5)}_0 - E^{(5)}_1\,,\label{E5mol}
\\
E^{(5)}_x =&\ -\frac{4}{3\,\pi}\,\frac{\mu_x}{m}\,\biggl(\ln k_0(\mathrm{H})+\ln\frac{\mu_x}{m}\biggr)
\nonumber \\ &\ 
+\frac{4}{3\,\pi}\,\biggl(\frac{\mu_x}{m}\biggr)^3
\biggl\{\biggl(1 + \frac{m}{4\,m_x} + \frac{m^2}{m_x^2}\biggr)\,\ln\bigl(\alpha^{-2}\bigr) + \frac{19}{30}
+ \frac{m}{m_x}\biggl(\frac{31}{6}+\frac{7}{2}\,\ln2\biggr)
      + \frac{m^2}{m_x^2}\,\ln\frac{m_x}{m}\biggr\},
\end{align}
\end{widetext}
where $\mu_x= m_x\,m/\left(m_x + m \right)$.
The Bethe logarithm is given by  \cite{Drake:90}
\begin{align}
\ln k_0 =&\ \frac{1}{\cal D}\,
\left\langle \vec J \,(H-E)\,\ln\bigl[2\,(H-E)\bigr]\,\vec J \right\rangle\label{08} 
\end{align}
where
\begin{align}
\vec J &=\frac{\vec p_0}{m_0} + \frac{\vec p_1}{m_1} - \frac{\vec p_2}{m} - \frac{\vec p_3}{m}\,, \\
{\cal D}&=\left\langle \vec J \,(H-E)\,\vec J \right\rangle 
= {\cal D}_0 + {\cal D}_1\,, \\ \label{Dx}
{\cal D}_x &=  \frac{2\,\pi}{\mu_x^2}\,\sum_{a}\langle \delta^3(r_{ax}) \rangle\,,
\end{align}
and the following numerical value of the atomic Bethe logarithm is used in the above
\begin{align}
\ln k_0(\mathrm{H})  =&\ 2.984\,128\,555\,765\,498\,.
\end{align}
In the formulas~(\ref{E5mol})-(\ref{Dx}) the expectation values are evaluated 
with the nonrelativistic wave function $\Psi$, and the notation in Eq. (\ref{E5mol}) 
$\langle \dots \rangle_\epsilon$ means the following limit
\begin{align}
\left\langle\frac{1}{r_{ij}^3}\right\rangle_{\!\!\epsilon} =
\lim_{\epsilon\rightarrow0}\left[\left\langle\frac{\theta(r_{ij}-\epsilon)}{r_{ij}^3}\right\rangle
+4\pi(\gamma+\ln \epsilon)\langle\delta^3(r_{ij})\rangle\right]\label{AS},
\end{align}
where the symbol $\gamma$ denotes the Euler-Mascheroni constant, and $\theta$ 
is the Heaviside function.

One subtle point to be clarified is the nuclear self-energy correction 
and the corresponding definition of the nuclear charge radius. 
This correction is insignificant for a regular hydrogen atom but 
non-negligible for muonic hydrogen ($\mu$H). So, for consistency with
the determination of the proton charge radius $r_p$ in $\mu$H \cite{pohl:10},
following Ref.~\cite{PKCP:19}, we account for this effect in the total energy of the hydrogen molecule
in a minimal way, by including in Eq.~(\ref{E5mol}) only logarithmic terms, and the nonlogarithmic terms
are absorbed into the mean square nuclear charge radius.

\section{Bethe logarithm}
Since the calculation of the Bethe logarithm $\ln k_0$
is the most complicated one, we describe below its evaluation in more detail,
extending our previous work \cite{PKCP:19} to two nuclei with different masses.
We express $\ln k_0$ in terms of the one-dimensional integral \cite{PK:04}
\begin{equation}
\label{lnk0ft}
\ln k_0 = \frac{1}{\cal D}\,\int_0^1 dt\, \frac{f(t)-f_0 - f_2\,t^2}{t^3}  
\end{equation}
with the function $f(t)$ defined as
\begin{equation}
f(t) = \biggl\langle\vec J\,\frac{k}{k+H-E}\,\vec J\biggr\rangle, \qquad t = \frac{1}{\sqrt{1+2\,k}} 
\label{ft}
\end{equation}
which has the following Taylor expansion
\begin{equation}
f(t) = f_0 + f_2\; t^2 + f_3\; t^3 + (f_{4l} \ln t + f_{4})\; t^4  + O(t^5)\label{ftexp}\,
\end{equation}
with the coefficients  {($m=1$)}
\begin{align}
  f_0 =&\ \langle J^2 \rangle\,, \nonumber \\
  f_2 =&\ -2\,{\cal D}\,, \nonumber \\
  f_3 =&\ \sum_x8\,\sqrt{\mu_x}\,{\cal D}_x\,, \nonumber \\
  f_{4l}=&\ \sum_x16\,\mu_x\,{\cal D}_x\,,  \\
  f_4 =&\ 4\,\bigg \langle \bigg[\sum_{a,x} \frac{1}{\mu_x}\,\frac{\vec r_{ax}}{r_{ax}^3}
 +\biggl(\frac{1}{m_0}-\frac{1}{m_1}\biggr)\,\frac{\vec r_{01}}{r_{01}^3}\bigg]^2\bigg \rangle_{\!\epsilon}\nonumber \\ &
   -2\,\sum_x{\cal D}_x\,\Big(1 + 4\,\mu_x \ln\frac{\mu_x}{4}   - 4\,\mu_x\Big) \,, 
\end{align}
which has been obtained from the known high-$k$ expansion by Korobov \cite{Korobov:12},
with all the terms proportional to $\delta^3(r_{01})$ being neglected.
The integrand in Eq.~\eqref{lnk0ft}, as a smooth function of $t$, was evaluated at 200 equally
spaced points in the range $t \in [0, 1]$, which enabled relative uncertainty higher than 
$10^{-7}$. In the numerical  calculation of $f(t)$, the resolvent in Eq. (\ref{ft}) was represented
in terms of pseudostates of the form  $\vec{\phi} ^{\,\Pi} =\vec{r}_{ab}\,\phi$ for all interparticle coordinates. 
The nonlinear parameters of $\vec{\phi} ^{\,\Pi}$ were found by a maximization of $f$.

The $f(1)$ value can be determined analytically
using the generalized Thomas-Reiche-Kuhn sum rule \cite{Zhou:06}
\begin{equation}
  \langle \vec J \,(H-E)^{-1}\,\vec J \rangle= \frac{3}{2}\,\bigg(\frac{m}{\mu_0}+\frac{m}{\mu_1}\bigg),
\end{equation}
which enables an assessment of the completeness of the pseudostates space 
and the uncertainty estimation.
For the given size $N$ of the wave function $\Psi$ expansion, the size of the pseudostate 
basis set was chosen as $N'=\frac{3}{2}N$, which appeared to be sufficient for most of the $t$ points. 
There were also additional factors taken into account for the accurate representation of
the resolvent in Eq. (\ref{ft}). 
The powers of the internuclear coordinate $r_{01}$, analogously to the wave function, 
were restricted to even integers and were generated randomly for each basis function 
from the log-normal distribution within the $0-80$ range. 
However, for small values of $t$ ($\leq 0.1$), 
due to a cancellation in the numerator of Eq.~(\ref{lnk0ft}), 
an additional tuning of the distribution was made and $N'=2\,N$ was set
to achieve high accuracy.
Moreover, in this critical region of small $t$, the function $f(t)$ was expanded in a power series
in Eq. (\ref{ftexp}), and the higher order expansion terms were obtained from the fit to numerical values of $f(t)$.
In order to perform the integration  in Eq.~\eqref{lnk0ft}, we used a polynomial interpolation 
of the integrand for $t > 0.1$, and a power expansion for the critical  region $t \in [0, 0.1]$.

The convergence of the Bethe logarithm with the increasing size
of the naECG basis is shown in Table~\ref{Tlnk}. Six significant figures can be considered
stable and the estimated relative uncertainty is a half ppm for all molecules,
as previously for H$_2$.
\begin{table*}[t!hb]
\caption{Convergence of the Bethe logarithm $\ln k_0$ 
  with the increasing size $N$ of the naECG basis set. The final uncertainty for $\ln k_0$
  is due to numerical inaccuracy of $f(t)$ at small $t$. }
\label{Tlnk}
\begin{ruledtabular}
\begin{tabular}{cx{3.11}x{3.11}x{3.11}x{3.11}x{3.11}}
\centt{$N$} & \centt{D$_2$} & \centt{T$_2$} & \centt{HD} & \centt{HT} & \centt{DT}\\
\hline\\[-1.5ex]
    128       & 3.016\,145\,65 &  3.016\,557\,24 & 3.018\,009\,62 & 3.018\,175\,20 &  3.017\,480\,26  \\
    256       & 3.018\,288\,11 &  3.018\,323\,28 & 3.018\,207\,59 & 3.018\,335\,11 &  3.018\,259\,13  \\
    512       & 3.018\,459\,13 &  3.018\,487\,17 & 3.018\,347\,02 & 3.018\,373\,25 &  3.018\,439\,11  \\
    1024      & 3.018\,473\,32 &  3.018\,514\,61 & 3.018\,385\,84 & 3.018\,414\,64 &  3.018\,484\,83  \\
    2048      & 3.018\,475\,98 &  3.018\,519\,89 & 3.018\,393\,11 & 3.018\,418\,33 &  3.018\,496\,12 \\
    $\infty$  & 3.018\,478(2)  &  3.018\,522(3)  & 3.018\,397(4)  & 3.018\,422(4)  &  3.018\,501(5)  \\
\end{tabular}                            
\end{ruledtabular}   
\end{table*}

\section{Higher order QED}

The higher order QED corrections are calculated within the Born-Oppenheimer approximation. 
First let us consider the second iteration of the relativistic correction ${\cal E}^{(4)}$
to the BO potential
\begin{equation}
E^{(6)}_\mathrm{sec}=\left\langle
\chi(R)\left|{\cal E}^{(4)}(R)\frac{1}{(E^{(2)}-H_{\rm n})'}{\cal E}^{(4)}(R)\right|\chi(R)
\right\rangle\,,
\end{equation}
where $\chi(R)$ is the radial nuclear wave function obtained from the radial Schr{\"o}dinger
equation with the Hamiltonian consisting of the nuclear kinetic energy and the
nonrelativistic BO potential.
This term is of $\alpha^6$ order and is considered separately for consistency
with the previous calculations of $E^{(4)}$ using the nonadiabatic perturbation theory.
The main $\alpha^6$ contribution is obtained by averaging the
${\cal E}^{(6)}(R)$ potential obtained in the BO framework in Ref. \cite{PKCP:16} 
and the unknown $1/\mu$ correction is estimated to be smaller than 
the numerical uncertainty of ${\cal E}^{(6)}(R)$. 

Because of the significant increase in the accuracy of the QED correction achieved in this work,
the dominating contribution to the uncertainty comes from the higher order $E^{(7)}$ correction.
Currently, an explicit form of this correction is unknown, which prevents its
accurate evaluation. Its first estimation, made within the BO approximation 
framework, was reported in Ref.~\cite{PKCP:16}. Here, following \cite{PKCP:19}, 
we account for several additional terms, namely
\begin{align}
  E^{(7)} \approx&\  \pi\,\Big\langle\sum_{a,x}\delta^3(r_{ax})\Big\rangle
  \bigg\{ \frac{1}{\pi}\big[ A_{60} + A_{61}\,\ln\alpha^{-2} \label{25} \\&\ 
  + A_{62}\, \ln^2\alpha^{-2}\big] + \frac{B_{50}}{\pi^2} + \frac{C_{40}}{\pi^3}\biggr\} - E^{(7)}_0 -E^{(7)}_1\,,
\nonumber
\end{align}
where $A,B$, and $C$ coefficients corresponds to the well  known one-~, two-, and three-loop hydrogenic Lamb shift \cite{Eides:01}.
Since we calculate the dissociation energy, the atomic values $E^{(7)}_{0,1}$ are subtracted out. 

Finally, at the achieved accuracy level, the nuclear finite size effect cannot be neglected
and it is accounted for by the following formula
\begin{align}
E^{(4)}_{\rm FS} =&\ \alpha^4\,
\frac{2\pi}{3}\,\Big\langle\sum_{a,x}\delta^3(r_{ax})\Big\rangle\,\frac{(r_{c0}^2+r_{c1}^2)}{2\,\lambdabar^2}
- E^{(4)}_{\rm FS0} - E^{(4)}_{\rm FS1}\,,\label{FS}
\end{align}
where $r^2_{c0/1}$ is the mean square charge radius of the nucleus $0/1$,
$\lambdabar$ is the electron Compton wavelength, and atomic values $E^{(4)}_{\rm FS0,1}$ are subtracted out.

\section{Final results and conclusion}

Theoretical predictions for the known contributions to the dissociation energies of all
molecular hydrogen isotopologues are presented in Table \ref{TD}.
\begin{table*}[!htb]
\renewcommand{\arraystretch}{1.2}
\caption{Theoretical predictions for the dissociation energy budget for
  the ground level of the molecular hydrogen isotopologues. 
  $E_\mathrm{FS}$ is the finite nuclear size correction with
  $r_p =  0.8414(19)$ fm \cite{CODATA_18}, $r_d = 2.12799(74)$ fm \cite{CODATA_18}, and $r_t =  1.7591(363)$ fm \cite{angeli_13}. 
  All the energy entries are given in $\icm$.}
\label{TD}
\begin{ruledtabular}
\begin{tabular}{lx{6.11}x{6.11}x{6.11}x{6.11}x{6.11}}
\centt{Contribution} & \centt{D$_2$} & \centt{T$_2$} & \centt{HD} & \centt{HT} & \centt{DT}\\    
\hline                                                            
$E^{(2)}$            & 36\,749.090\,990\,00(1) & 37\,029.224\,867\,00(1) & 36\,406.510\,890\,07(1) & 36\,512.928\,009\,11(1) & 36\,882.009\,843\,48(1) \\
$E^{(4)}$            &      -0.528\,206\,05(9) &      -0.526\,750\,0(2)  &      -0.529\,887\,5(2)  &      -0.529\,377\,9(2)  &      -0.527\,523\,6(3)  \\
$E^{(5)}$            &      -0.198\,256(3)     &      -0.199\,735(4)     &      -0.196\,441(4)     &      -0.197\,005(5)     &      -0.198\,958(5)     \\
$E^{(6)}$            &      -0.002\,096(6)     &      -0.002\,110(6)     &      -0.002\,080(6)     &      -0.002\,085(6)     &      -0.002\,103(6)     \\
$E^{(6)}_{\rm sec}$  &       0.000\,009\,4     &       0.000\,009\,4     &       0.000\,009\,3     &       0.000\,009\,3     &       0.000\,009\,4     \\
$E^{(7)}$            &       0.000\,103(25)    &       0.000\,103(25)    &       0.000\,102(25)    &       0.000\,102(25)    &       0.000\,103(25)    \\
$E^{(4)}_\mathrm{FS}$&      -0.000\,202        &      -0.000\,139(6)     &      -0.000\,116        &      -0.000\,084(3)     &      -0.000\,171(3)     \\
Total                & 36\,748.362\,342(26)    &  37\,028.496\,245(27)   & 36\,405.782\,477(26)    & 36\,512.199\,569(26)    & 36\,881.281\,200(26)    \\[1ex] 
Exp.	               & 36\,748.362\,86(68)     &                         & 36\,405.783\,66(36)     &   &        \\
Diff.                &       0.000\,52(68)     &                         &       0.001\,18(36) \\
\end{tabular}                            	
\end{ruledtabular}   
\end{table*}
Thanks to the direct nonadiabatic calculation
of  the nonrelativistic energy \cite{PK:18a,PK:18b,PK:19}
and also of the relativistic \cite{PSKP:18,Wang:18a,Wang:18b} and leading quantum electrodynamic corrections,
the theoretical dissociation energy of all the isotopologues of molecular hydrogen
has reached the level of 0.8~MHz ($26\times 10^{-6}\,\icm$ or $8\times 10^{-10}$ of relative uncertainty).
The higher order $m\,\alpha^6$ QED contribution has been calculated \cite{PKCP:16} within the BO approximation,
but the corresponding uncertainty is almost negligible.
At present the accuracy of theoretical predictions is limited by the poorly known $E^{(7)}$ term
of the $\alpha$-expansion~(\ref{alphaexp}), which has been estimated using
the atomic hydrogen values with 25\% uncertainty, as in Ref. \cite{PKCP:19}.
As a result, the significantly improved theoretical predictions for the ground state dissociation energy 
of the D$_2$ molecule (as well as for H$_2$) are in very good agreement with the most recent measurement 
\cite{Liu:09}, but the experimental uncertainty is more than 20 times larger than the theoretical one.
The situation is more intriguing for the dissociation energy of the HD molecule.
Our theoretical prediction differs by 3$\,\sigma$
from the most recent measurement in Ref.~\cite{Sprecher:10}. If this experimental value is confirmed,
this could indicate the existance of yet unknown physical effects, which are
specific to heteronuclear molecules only.

{\sl Acknowledgments} --
We thank referees for their pertinent suggestions which helped to improve the paper.
This research was supported by National Science Center (Poland) Grants No. 2014/15/B/ST4/05022 (M.P.)
and 2017/25/B/ST4/01024 (J.K.) as well as by a computing grant from 
Pozna\'n Supercomputing and Networking Center and by PL-Grid Infrastructure.


\appendix
\section{Expectation values of individual operators}

In Table V we present nonrelativistic energies and expectation values (in a.u.) of
individual operators with the nonrelativistic 4-body naECG wave function for a possible comparison with any future calculations.
  
\begin{table*}[!ht]
\renewcommand{\arraystretch}{1.4}
\caption{Mean values of various operators with naECG wave function for the ground molecular state}
\begin{ruledtabular}
\begin{tabular}{cx{2.15}x{2.15}x{3.15}x{2.15}x{3.14}}
\centt{Operator} & \centt{D$_2$} & \centt{T$_2$} & \centt{HD} & \centt{HT} & \centt{DT} \\
\hline\\[-1.5ex]
H                        & -1.167\,168\,809\,26(4)        &  -1.168\,535\,675\,59(17) & -1.165\,471\,923\,93(6) &  -1.166\,002\,037\,24(12) &  -1.167\,819\,673\,31(16)      \\
$\vec J^2$               &  2.527\,565\,218\,16(15)       &   2.531\,714\,163\,5(6)   &  2.522\,506\,446\,2(3)  &   2.524\,076\,110\,3(5)   &   2.529\,532\,438\,3(8)       \\
$\sum_a 4\,\pi\,\delta(r_{0a})$        
                         &  5.703\,646\,95(5)             &   5.716\,698\,5(2)        &  5.685\,102\,31(6)      &   5.689\,493\,73(8)       &   5.709\,140\,6(3)            \\
$\sum_a 4\,\pi\,\delta(r_{1a})$        
                         &  5.703\,646\,95(5)             &   5.716\,698\,5(2)        &  5.689\,495\,06(7)      &   5.695\,351\,91(11)      &   5.710\,607\,5(4)            \\
$4\,\pi\,\delta(r_{23}) $        
                         &  0.205\,013\,236(5)            &   0.205\,964\,94(3)       &  0.203\,833\,913(15)    &   0.204\,202\,05(2)       &   0.205\,466\,14(3)           \\
$\sum_a p_a^4$           &-13.076\,757\,2(3)              & -13.106\,197\,7(9)        & 13.039\,562\,0(3)       & -13.051\,261\,4(5)        & -13.090\,834\,4(7)             \\
$p_2^i \Big( \frac{\delta^{ij}}{r_{23}} + 
\frac{r_{23}^i r_{23}^j}{r_{23}^{3}} \Big) p_3^j$ 
                        & 0.093\,386\,335(4)              &  0.093\,728\,50(2)        &  0.092\,959\,18(4)      &    0.093\,093\,113(14)    & 0.093\,549\,582(9)             \\
$\sum_a p_0^i\,\Big(\frac{\delta^{ij}}{r_{0a}}+ 
\frac{r_{0a}^i\,r_{0a}^j}{r_{0a}^3}\Big)\,p_a^j $           
                        &  -2.507\,167\,22(8)             &  -2.512\,657\,3(3)        & -2.495\,702\,4(3)       & -2.496\,033\,9(4)         & -2.507\,855\,6(5)              \\
$\sum_a p_1^i\,\Big(\frac{\delta^{ij}}{r_{1a}}+
\frac{r_{1a}^i\,r_{1a}^j}{r_{1a}^3}\Big)\,p_a^j $           
                        &  -2.507\,167\,22(8)             &  -2.512\,657\,3(3)        & -2.504\,864\,8(3)       & -2.508\,852\,1(4)         & -2.511\,717\,0(5)              \\
$p_0^i\,\Big(\frac{\delta^{ij}}{r_{01}}+
\frac{r_{01}^i\,r_{01}^j}{r_{01}^3}\Big)\,p_1^j $         
                        & -17.802\,138\,7(4)              & -21.964\,332\,6(11)       & -14.388\,033\,5(7)      & -15.304\,814\,1(15)       & -19.550\,968\,3(15)            \\
$\sum_a \langle r_{0a}^{-3}\rangle_\epsilon$ & -3.614\,687\,9(8)               & -3.622\,724(3)            & -3.602\,242\,6(8)       &   -3.604\,722\,9(11)      & -3.617\,784(2) \\
$\sum_a \langle r_{1a}^{-3}\rangle_\epsilon$ & -3.614\,687\,9(8)               & -3.622\,724(3)            & -3.606\,688\,1(7)       &   -3.610\,649\,1(8)       & -3.619\,274(3) \\
$\langle r_{23}^{-3}\rangle_\epsilon$        &  0.405\,522\,77(4)              &  0.407\,090\,24(16)       &  0.403\,586\,4(5)       &    0.404\,190\,4(11)      &  0.406\,268\,9(9)\\
$\sum_a \langle r_{0a}^{-4}\rangle_\epsilon$ & -2.865\,27(2)                   & -2.869\,32(5)             & -2.871\,97(3)           &   -2.877\,83(4)           & -2.895\,29(4)   \\
$\sum_a \langle r_{1a}^{-4}\rangle_\epsilon$ & -2.865\,27(2)                   & -2.869\,32(5)             & -2.875\,51(3)           &   -2.882\,62(4)           & -2.896\,50(4)  \\
$\sum_{a<b}\frac{\vec r_{0a}}{r_{0a}^3} \cdot \frac{\vec r_{0b}}{r_{0b}^3} $           
                        & -0.011\,661\,1(7)               & -0.011\,686(3)            & -0.011\,580\,5(16)     &    -0.011\,571\,3(5)       & -0.011\,653\,9(4)              \\
$\sum_{a<b}\frac{\vec r_{0a}}{r_{0a}^3} \cdot \frac{\vec r_{1b}}{r_{1b}^3} $           
                        & -0.234\,951\,712(6)             & -0.236\,540\,76(3)        & -0.232\,986\,687(5)    &    -0.233\,599\,584(7)     & -0.235\,707\,73(3)             \\
$\sum_{a<b}\frac{\vec r_{1a}}{r_{1a}^3} \cdot \frac{\vec r_{1b}}{r_{1b}^3} $           
                        & -0.011\,661\,1(6)               & -0.011\,686(3)            & -0.011\,677\,8(2)      &    -0.011\,704\,7(5)       & -0.011\,687\,7(4)              \\
$\sum_a\frac{\vec r_{01}}{r_{01}^3} \cdot \frac{\vec r_{0a}}{r_{0a}^3} $           
                        & -1.173\,53(6)                   & -1.177\,4(2)              & -1.167\,84(9)          &    -1.168\,92(12)          &  -1.174\,56(10)                \\
$\sum_a\frac{\vec r_{01}}{r_{01}^3} \cdot \frac{\vec r_{1a}}{r_{1a}^3} $          
                        & 1.173\,53(6)                    &  1.177\,4(2)              &  1.169\,50(8)          &     1.171\,16(11)          &  1.175\,13(11)                 \\[2ex]
\end{tabular}                            
\end{ruledtabular}   
  \end{table*}

\end{document}